\documentstyle[12pt,psfig]{article}
\def\4v#1{{\rm #1}}

\def\3v#1{{\bf #1}}

\def\fslash#1{\not\! {\4v #1}}

\newcommand{\beq}{\begin{equation}}
\newcommand{\eeq}{\end{equation}}
\newcommand{\Imag}{\mbox{Imag}}
\newcommand{\Real}{\mbox{Real}}
\newcommand{\be}{\begin{equation}}
\newcommand{\ee}{\end{equation}}
\newcommand{\bea}{\begin{eqnarray}}
\newcommand{\eea}{\end{eqnarray}}
\begin{document}
\title{On the Dirac Structure of the Nucleon Selfenergy\\
in Nuclear Matter}

\author{A. Trasobares, A. Polls, A. Ramos\\
\em Departament d'Estructura i Constituents de la Mat\`eria\\
\em Universitat de Barcelona,
E-08028 Barcelona, Spain\\
H. M\"uther\\
\em Institut f\"{u}r Theoretische Physik,\\ \em Universit\"{a}t
T\"{u}bingen,
D-72076 T\"{u}bingen, Germany}
\date{}

\maketitle

\begin{abstract}
The relativistic structure of the self-energy of a nucleon in nuclear matter 
is investigated including
the imaginary and real components which arise from the terms of first and second
order in the NN interaction. A parameterized form of Brueckner $G$ matrix is
used for the NN interaction. The effects of the terms beyond the DBHF
approximation on quasiparticle energies and the optical potential for
nucleon-nucleus scattering are discussed.
\end{abstract}

\section{Introduction}

During the last few years the attempts to derive the ground-state properties 
of nuclear systems from a realistic nucleon-nucleon (NN) interaction, have been
promoted very much by the understanding that relativistic effects may be
non-negligible in such investigations. These ideas were originally developed
within the various versions of the phenomenological Walecka model\cite{Walecka}.
The Dirac structure of the NN interaction with a strong repulsive component 
originating from the exchange of the $\omega$ vector meson and an attractive
component of medium range described in terms of a scalar meson ($\sigma$) leads
to a self-energy of the nucleon in the nuclear medium which contains a large
scalar component $\Sigma^s$ and a large time like vector component $\Sigma^0$. 
These two components
compensate each other to a large extent if one calculates the single-particle
energy. This leads to the well known fact that the binding-energy of nuclei are
very small as compared to the rest mass of the nucleon M, a fact which has often
been used to argue that relativistic effects should be small in the many-body
problem of nuclear physics. The individual components of the self-energy
$\Sigma^s$ are of the order of the nucleon rest mass (typically one third of
$M$). Therefore the structure of the Dirac spinors in the nuclear medium is 
modified to quite some extent as compared to the Dirac spinor for a free
nucleon. This medium dependence of the Dirac spinors, which affects the
evaluation of the NN interaction in the medium, leads to a saturation mechanism
for nuclear matter.

These more or less empirical models received support from calculations, which
start from a realistic One-Boson-Exchange model of the NN 
interaction\cite{rupr}. In this context realistic NN interaction means that the
parameters contained in these models are adjusted to describe the experimental
data of free NN scattering. Using such a realistic NN interaction in a many-body
calculation of the Dirac-Brueckner-Hartree-Fock (DBHF) type, results were
obtained for the saturation point of nuclear matter which were in quite a good
agreement with the empirical saturation point\cite{brock,malf}. 
Such a DBHF calculation accounts
for the effect of correlations on the level of the BHF approximation, i.e.~the
lowest order in the hole line expansion, and allows for the relativistic effects
which we just outlined. This success of the relativistic features contained in
the DBHF approach could be a solution of an old problem: the so-called Coester
band phenomenon\cite{coester}, which stands for the fact that many-body
calculations based on various realistic models of the NN interaction lead to
predictions for the saturation point of nuclear matter, which either fail to
yield enough binding or predict a saturation density twice as large as the
empirical value.

Such a Coester band can also be observed in non-relativistic studies of finite
nuc\-lei\cite{coes1}. Therefore it was quite obvious that attempts have 
been made
to include the relativistic features also in DBHF calculations of finite nuclei.
Indeed the inclusion of relativistic effects gave a substantial improvement in
the calculated binding energies and radii of finite nuclei\cite{fritz,boersma}. 
However, there remains still a discrepancy between the DBHF results and the
experimental values. 

On the other hand it was observed that an extension of the non-relativistic BHF
approach to a definition of the nucleon self-energy which accounts in a
symmetric way not only for the particle-particle ladders included in the
Brueckner $G$-matrix but also for the corresponding hole-hole scattering term
may lead to a larger binding energy and a larger radius than obtained in the BHF
approach\cite{skour}. A similar feature, moving the saturation point away from
the Coester band by including hole-hole scattering terms, has also been observed
in particle-particle hole-hole RPA calculation of nuclear 
matter\cite{kuo1,kuo2}. Taking a very optimistic point of view one 
may argue that the combination
of relativistic effects and hole-hole scattering terms may lead to an improved
microscopic understanding of groundstate properties of nuclear systems.

In the work presented here we would like to investigate the effects of
particle-particle and hole-hole scattering terms on the relativistic structure
of the nucleon self-energy. For that purpose we consider the nucleon self-energy
defined with all terms of first and
second order in the NN interaction (see Fig.~1). In contrast to earlier
investigations\cite{Horowitz,Chin,Steffani,ylhan} we evaluate the imaginary
contributions to the nucleon self-energy allowing for all possible 
combinations of momentum
$\3v k$  and energy $k^0$. Therefore we can use dispersion relations to evaluate
also the real part of the second order diagrams. We investigate
in detail the effect of these higher order diagrams on the Dirac structure of
the self-energy. Furthermore we derive from this relativistic
self-energy an optical potential to be used in a Schr\"odinger equation for
nucleon-nucleus scattering.

The technical details for the
calculation of the imaginary components in the self-energy are presented
in section 2. The
dispersion relations used to evaluate the corresponding real components are 
given in section 3. That section also contains a detailed discussion of the
results. The main conclusions are summarized in the final section.

\begin{figure}
\psfig{figure=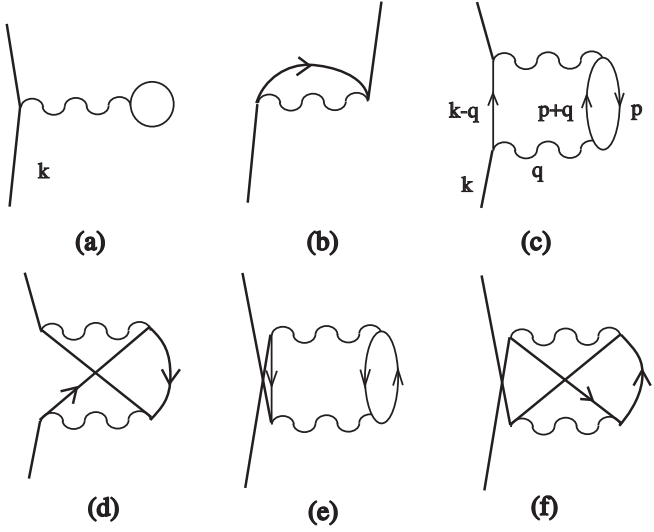,height=70mm}
\caption{Graphical representation of the Hartree (a), Fock (b), 
2p1h direct (c), 2p1h exchange (d), 2h1p direct (e) and 2h1p exchange (f)
contributions to the self-energy of the nucleon. 
The momenta labelling the
various contractions in (c) correspond to the nomenclature used in 
(\protect\ref{eq:sigmax} )
}
\end{figure}

\section{Imaginary part of the nucleon self-energy}

The physical system of neutron or nuclear matter is, by definition,
translational as well as rotational invariant. Furthermore it is also
assumed to be invariant under parity and time reversal. In general,
under these conditions, the nucleon self-energy can be
written\footnote{Notation: within formul\ae \ we shall
use roman type (${\rm p,q,\cdots}$) for 4-vectors, boldface type
(${\bf p,q,\cdots}$) for 3-vectors and normal math italic
type (${p,q,\cdots}$) for the norm of 3-vectors.} as \cite{Walecka}:
\beq
\Sigma(\4v k)=\Sigma^s(\4v k)-\gamma_{0}
  \Sigma^0(\4v k)+
  \mbox{\boldmath$\gamma$}\cdot{\3v k} \Sigma^v(\4v k)
\label{eq:Sigma}
\eeq
where the functions $\Sigma^r$ ($r=s,0,v$) can be projected
out by taking the appropriate traces:
\begin{eqnarray}
\Sigma^s&=&{1\over 4} Tr\lbrack\Sigma\rbrack \\
\Sigma^0&=&{-1\over 4} Tr\lbrack\gamma_{0}\Sigma\rbrack \\
\Sigma^v&=&{-1\over {4{\3v k}^2}}
Tr\lbrack \mbox{\boldmath$\gamma$}\cdot{\3v k} \ \Sigma \rbrack
\end{eqnarray}

As mentioned in the introduction, in this work we want to go beyond 
the usual Hartree-Fock approximation to the
self-energy, visualized by the diagrams a and b of Fig.~1, and pay
special attention to the 2hole-1particle (2h1p) and 2particle-1hole (2p1h)
diagrams displayed in Figs. 1c, 1d, 1e and 1f.
Therefore, the different components of the self-energy are complex
functions and, in general, we will write them as:
\beq
\Sigma^r=V^r+iW^r
\eeq
%

Previous works have focused on the imaginary part of the self-energy at
the on-shell energy of the propagating nucleon. In Ref. \cite{Horowitz}
the 2p1h and 2h1p direct contributions where studied, while the work of Ref.
\cite{Steffani} considered also the Fock-exchange terms. This study, however,
was  restricted to the on-shell 2h1p contributions.

In the present work we extend these
calculations and evaluate the direct and exchange terms for both 2h1p
and 2p1h contributions to the nucleon self-energy considering off-shell effects,
i.e.~investigating the self-energy $\Sigma(k_0, \3v k)$ for all combinations of
energy $k_0$ and momentum $\3v k$. In a first step we
calculate the imaginary part of the self-energy, as
described below, from which we later obtain the real part by means of a
dispersion relation.

The nucleon-nucleon interaction is derived from a $G$-matrix evaluated 
within the Dirac-Brueckner-Hartree-Fock approach. This $G$-matrix is
parameterized in terms of an exchange
of effective $\sigma$ and $\omega$ mesons\cite{fritz,Muether}. The
starting point
of our discussion will be the Hartree-Fock (HF) approach.
Therefore, and for the sake of simplifying the notation, we will
define the following effective quantities in terms of the Hartree-Fock
components of the self-energy:
\begin{eqnarray}
M^* (\3v k ) &=& M + \Sigma^s_{HF} (\3v k)  \label{eq:s6}\\
{\3v k}^* &=& {\3v k} ( 1 + \Sigma^v_{HF} (\3v k)) \\
E^* (\3v k ) &=& \sqrt{{\3v k}^{*2} + M^{*2}}\label{eq:s8} \\
{\4v k}^* &\equiv& (k_0^*, {\3v k}^* ) = ( k_0 + \Sigma^0_{HF} (\3v k), 
{\3v k}^*)\\
\epsilon_{HF} (\3v k) = E^* (\3v k ) - \Sigma^0_{HF} (\3v k)\label{eq:epshf}
\end{eqnarray}
The nucleon HF propagator is then
\beq
G (\4v k ) =  \frac{{\fslash k}^* + M^*(\3v k)}{2 E^*(\3v k)} g(\4v k ) 
\label{eq:greenhf}
\eeq
where
\beq
g (\4v k ) = \frac{\theta ( E^*(\3v k) - E_F^* )}{k_0^* - E^*(\3v k) + i\eta}
+ \frac{\theta ( E_F^* - E^*(\3v k) )}{k_0^* - E^*(\3v k) - i\eta} -
\frac{1}{k_0^* + E^*(\3v k) - i\eta}
\label{eq:prop2}
\eeq
with $E_F^* = \sqrt{k_F^{*2} + M^{*2}}$. In eq. (\ref{eq:prop2})
we will disregard the last term, which is the anti-nucleon
contribution. This is quite reasonable
due to two reasons. First of all, the production of fermion
anti-fermion pairs should be negligible at the energy scales
of interest and secondly the pair production
 is inhibited in high density matter \cite{Horowitz,Chin}.
In the actual calculations we  set $\Sigma^v_{HF} = 0$ and disregard
the momentum dependence of $\Sigma^s_{HF}$ and
$\Sigma^0_{HF}$ for which we take their values at the Fermi momentum
$k_F$. This approximation relies on the weak momentum dependence
of the Hartree-Fock self-energies \cite{Walecka}. In this way,
$M^*$ acquires a fixed value, ${\3v k}^* = {\3v k}$, and the
only effective quantity that depends on the nucleon momentum
is $E^*_{\4v k}$, which from now on will be labelled with a subindex.
It is also useful to split up the fermion-propagator into
a particle and a hole term  and consider explicitly all
terms of Fig.~1 by taking the following prescriptions for
the nucleon propagators in the diagrams:
\begin{eqnarray}
{\rm Particle \ line} &:& iG_p (\4v k ) = i{{{\fslash k}^*
  +M^{*}}\over{2E^{*}_{\4v k}}} \ g_{p}(\4v k) 
\label{eq:part}\\
  {\rm Hole \ line} &:& iG_h (\4v k ) = i{{{\fslash k}^*
  +M^{*}}\over{2E^{*}_{\4v k}}} \ g_{h}(\4v k) \ ,
\label{eq:hole}
\end{eqnarray}
where the $g$-functions for particle and hole states are defined as:
$$g_{p}({\rm k}) = {{\theta(E^*_{\4v k} -E_{F})}\over{k_{0}^* - E_{\4v
k}^*
  +i\eta}}$$
$$g_{h}({\rm k}) = {{\theta(E_{F}^*-E^*_{\4v k})}\over{k_{0}^* - E_{\4v
k}^*
  -i\eta}}$$
These $g$-functions are formally similar to
the non-relativistic HF
nucleon propagators.
Therefore, applying the standard Feynman rules together with the
prescriptions of Eqs. (\ref{eq:part}) and (\ref{eq:hole}), 
we obtain a general expression for
the 2p1h contribution
\beq
i\Sigma_{X;ab}(k_{0},k)=i
  \int {d^4{\rm q}\over (2\pi)^4}{{d^4{\rm p}^*}\over (2\pi)^4}
  {\cal M}_{X;ab}({\rm k,p,q}) g_{p}({\rm p+q})g_{h}({\rm p})
   g_{p}({\rm k-q}) \label{eq:sigmax}
\eeq
while the formula for the 2h1p state is identical to the above except for
the exchange of the $p$ and $h$ subscripts in the $g$-functions. The
subscript $X$ stands for $D$ (direct) and $E$ (exchange) diagrams, while
$a$ and $b$ stand for the different meson types.

The function ${\cal M}_{X;ab}({\rm k,p,q})$ contains, basically, the Dirac
structure of the interaction and, for direct terms, it has the general form,
\begin{eqnarray}
{\cal M}_{D;ab}({\rm k,p,q})={{-\lambda_{I} \ g_{a}^2 \ g_{b}^2}
  \over{8E_{\4v p}^* E_{\4v p+q}^* E_{\4v k-q}^* }} \
  \Gamma_{b}({\rm q}) \
  \Delta_{b}({\rm q}) \ Tr\lbrack \Gamma_{b}({\rm q}) \
  (\fslash p^* +\fslash q^* +M^{*}) \
  \Gamma_{a}({\rm q}) \nonumber \\
  (\fslash p^* +M^{*}) \rbrack
  (\fslash k^*  - \fslash q^* +M^{*}) \
  \Delta_{a}({\rm q}) \
  \Gamma_{a}({\rm q})
\label{eq:Md}
\end{eqnarray}
The constant $\lambda_{I}$ is the isospin degeneracy and comes from
the loop trace, $g_{a}$ and $g_{b}$ are the meson-nucleon coupling constants.
The meson-nucleon vertices are denoted by $\Gamma$ and meson
propagators by $\Delta$. A form factor of the type $F({\4v q})=
\frac{\Lambda^2}{\Lambda^2 - {\4v q}^2}$, with a typical cut-off
mass of $\Lambda = 1500$ MeV, has been attached to each
vertex. This is equivalent to modifying the meson propagators
in the following way:
\beq
\Delta_a ({\4v q}) \longrightarrow \Delta_a ({\4v q})
F^2 ({\4v q})
\eeq
Note as well the minus sign that comes
from the fermion loop. For exchange terms one can write:
\begin{eqnarray}
{\cal M}_{E;ab}({\rm k,p,q})={{g_{a}^2 \ g_{b}^2}
  \over{8E_{\4v p}^* E_{\4v p+q}^* E_{\4v k-q}^* }} \
  \Gamma_{b}({\rm k}-{\rm q}-{\rm p}) \
  \Delta_{b}({\rm k}-{\rm q}-{\rm p}) \
  (\fslash p^* +\fslash q^* +M^{*}) \nonumber \\
  \Gamma_{a}({\rm q}) \
  (\fslash p^* +M^{*}) \
  \Gamma_{b}({\rm k} -{\rm q} -{\rm p}) \
  (\fslash k^*  - \fslash q^* +M^{*}) \
  \Delta_{a}({\rm q}) \
  \Gamma_{a}({\rm q})
\label{eq:Me}
\end{eqnarray}

The imaginary part of the self-energy is obtained as:
\beq
W_{X;ab}(k_{0},k)=4\pi^3 \int {d^4{\rm q}\over (2\pi)^4}{{d^4{\rm p}^*}
  \over (2\pi)^4}
  {\cal M}_{X;ab}({\rm k,p,q}) \ \Theta({\rm k,p,q})
\eeq
Where, for 2p1h, the factor $\Theta$ has the following form:
\begin{eqnarray}
\Theta({\rm k,p,q})=\theta(E_{F}^*-p_{0}^*)
  \theta(k_{0}^*-q_{0}-E_{F}^*)
  \theta(p_{0}^*+q_{0}-E_{F}^*) \nonumber \\
  \delta(p_{0}^*-E_{\4v p}^*)
  \delta(p_{0}^*+q_{0}-E_{\rm p+q}^*)
  \delta(k_{0}^*-q_{0}-E_{\rm k-q}^*)
\end{eqnarray}
and, for 2h1p:
\begin{eqnarray}
\Theta({\rm k,p,q})=-\theta(p_{0}^*-E_{F}^*)
  \theta(E_{F}^*-k_{0}^*+q_{0})
  \theta(E_{F}^*-p_{0}^*-q_{0}) \nonumber \\
  \delta(p_{0}^*-E_{\4v p}^*)
  \delta(p_{0}^*+q_{0}-E_{\rm p+q}^*)
  \delta(k_{0}^*-q_{0}-E_{\rm k-q}^*)
\end{eqnarray}
One can now use the step functions to define the boundaries for
the integrations
over $dq_{0}$ and $dp_{0}$. Working in spherical coordinates we can
 automatically perform one of the axial integrations (say over the
axial {\bf q}-angle).
By rewriting the first $\delta$-function in terms of the
$3$-momentum variable, the $dp$ integration can be easily performed.
 After that we obtain, for the 2p1h diagrams,
\begin{eqnarray}
W_{X;ab}(k_{0},k)={1\over{2(2\pi)^4}}
  \int\limits_{0}^{k_{0}^*-E_{F}^*} dq_{0} 
\int dq \ q^2 \ 
d(\cos\theta_{q}) \
\int\limits_{\bar p_{0}}^{E_{F}^*}
  dp_{0}^* \  p \int
  d\varphi_{p} \ 
  d(\cos\theta_{p}) \nonumber \\
  \quad {\cal M}_{X;ab}({\rm k,p,q}) \ E_{\4v p}^* \
  \delta(p_{0}^*+q_{0}-E_{\4v p+q}^*)
  \delta(k_{0}^*-q_{0}-E_{\4v k-q}^*)
\label{eq:2p1h}
\end{eqnarray}
Where, $p$ takes the value
$ p = \sqrt{p_{0}^{*2}-M^{*2}}$, and
\beq
\bar p_{0}=max(E_{F}^*-q_{0},M^{*}).
\eeq
For the 2h1p contribution  one obtains a similar
expression up to an overall minus sign.
The polar integration over the angle $\theta_q$ between ${\3v q}$ and the
external momentum ${\3v k}$ (taken along the z-axis) is readily
performed by using the delta function
$$\delta(k_{0}^*-q_{0}-E_{\bf k-q}^*) \equiv {E_{\4v k-q}^*\over{kq}} \
  \delta(\cos\theta_{kq}-{{2k_{0}^*q_{0}-{\rm q}^2+M^{*^2}-{\rm k}^{*2}}
  \over{2 k q}}) \ ,$$
which,
in order to keep $\mid \cos{\theta_{kq}} \mid < 1$,
imposes the following constraints to the $q$ variable:
$$q_{-}\leq q \leq q_{+}$$
where:
$$q_{\pm} = \vert p \pm \sqrt{(k_{0}^*-q_{0})^2-M^{*^2}} \vert \ .$$

The integration over the polar angle $\theta_p$ 
can also be performed very easily by referring
the two
angular variables of the momentum ${\3v p}$ ($\phi_p$,$\theta_p$)
to a reference frame 
in which ${\3v q}$ acts as the z-axis and using the remaining
$\delta$-function
$$\delta(p_{0}^*+q_{0}-E_{\4v p+q}^*) \equiv {E_{\4v p+q}^*\over{ p q}} \
\delta(\cos\theta_{pq}-{{2p_{0}^*q_{0}+{\rm q}^2}\over{2 p q}}) \ .$$
The requirement that the absolute value of the cosine must be 
less than one puts restrictions to the values of $p^{*}_0$, similar
to those obtained for $q$, but more complicated to be implemented 
analytically due to the two different sign possibilities for q$^2$.
For this reason, we have taken care of the restrictions over $p^*_0$
numerically through an explicit step function inside the integral.

After all these considerations we can write our final expression
for the imaginary part of the self-energy. For the 2p1h state
we have:
\begin{eqnarray}
W_{X;ab}(k_{0},k)={1\over{2(2\pi)^4k}}
\int\limits_{0}^{k_{0}^*-E_{F}^*} dq_{0}
\int\limits_{q_{-}}^{q_{+}} dq
\int\limits_{\bar
p_{0}}^{E_{F}^*} dp_{0}^* \  
\theta ( 1 - \left[ \frac{2p_0^* q_0 + {\4v q}^2}
{2 p q} \right]^2 )
\int\limits_{0}^{2\pi}
d\varphi_{p}  \nonumber \\
\tilde{\cal M}_{X;ab}(k_{0}^*,k,p_{0}^*,q_{0},q,\varphi_{p}) ,
\end{eqnarray}
and for the 2h1p state:
\begin{eqnarray}
W_{X;ab}(k_{0},k)={-1\over{2(2\pi)^4k}}
\int\limits_{k_{0}^*-E_{F}^*}^{0} dq_{0}
\int\limits_{q_{-}}^{q_{+}} dq
\int\limits_{E_{F}^*}^{E_{F}^*-q_{0}}
dp_{0}^* \ 
\theta ( 1 - \left[ \frac{2p_0^* q_0 + {\4v q}^2}
{2 p q} \right]^2 )
\int\limits_{0}^{2\pi}
d\varphi_{p}  \nonumber \\
\tilde{\cal M}_{X;ab}(k_{0}^*,k,p_{0}^*,q_{0},q,\varphi_{p}) \ ,
\end{eqnarray}
where it is understood that $\tilde{\cal M}$ does not contain
the energies in the denominators of Eqs. (\ref{eq:Md}) and
(\ref{eq:Me}) becasue they have canceled out
with the energy factors that appeared on rewriting
the $\delta$-functions.

\begin{figure}[tb]
\psfig{figure=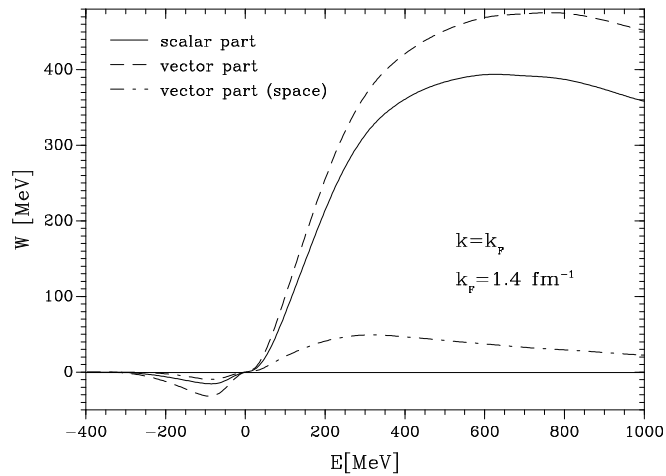,width=9cm}
\caption{Imaginary part of the self-energy calculated for a fixed momentum $k$ =
$k_F$ = 1.4 fm$^{-1}$ as a function of the energy variable, which is
normalized such that it is zero at the Fermi energy. Results are displayed for
the scalar part ($W^s$, solid line), the time-like vector part ($W^0$, dashed
line) and the space-like vector component ($W^v$, dashed dotted line)}
\end{figure}

\section{Results and Discussion}

In the first part of this section we will concentrate the discussion of results
on the example of nuclear matter at a density close to the saturation density,
in particular we will consider a Fermi momentum $k_F$ of 1.4 fm$^{-1}$. We are
going to discuss the contribution of the diagrams of second order in the NN
interaction to the various components of the imaginary part in the nucleon
self-energy. The NN interaction is described in terms of the exchange of an 
effective scalar, $\sigma$, meson and an effective vector, $\omega$, meson. The
masses of these mesons are fixed to the masses of the corresponding mesons in a
realistic meson exchange model of the NN interaction ($m_{\sigma}$ = 550 MeV, 
$m_{\omega}$ = 782.6 MeV), while the effective coupling constants have been
adjusted \cite{fritz} 
such that a Dirac-Hartree-Fock calculation of nuclear matter at $k_F$
= 1.4 fm$^{-1}$ using these constants would reproduce the results for
single-particle energies and the binding energy obtained in a 
Dirac-Brueckner-Hartree-Fock (DBHF) calculation employing Bonn A NN
potential\cite{rupr}. This Dirac-Hartree-Fock parameterization of the DBHF
results yields at this density coupling constants of $g_{\sigma}$=8.536 and 
$g_{\omega}$=9.536 for the
$\sigma$ and the $\omega$ meson, respectively\cite{fritz}. The single-particle
Green function $G$ of eq.(\ref{eq:greenhf}) for the propagation of the
intermediate states is also defined employing the results of this DBHF
calculation. This means that we consider a HF self-energy defined in terms of 
\begin{eqnarray}
\Sigma_{HF}^s & = & - 374.9\quad {\mbox{MeV}}\nonumber\\
\Sigma_{HF}^0 & = & - 289.8\quad {\mbox{MeV}} \, .\label{eq:selfhf}
\end{eqnarray}

As a first example we show in Fig.~2 the imaginary part of the self-energy
calculated for nucleons with a momentum fixed to $k$ = $k_{F}$ = 1.4 fm$^{-1}$ 
as a function of the energy. Note that the energy variable in this figure
is normalized such that an energy zero corresponds to the Fermi energy 
$\epsilon_F = \epsilon_{HF} (k_F)$ as defined in (\ref{eq:epshf}). 
Results are displayed for the
scalar component, $\Imag{\Sigma^s}=W^s$, the time-like vector component,
$\Imag{\Sigma^0}=W^0$, and the space-like vector component,
$k\Imag{\Sigma^v}=W^v$. All these components are negative at energies below
the Fermi energy and positive for those above. They are much larger at
positive energies, reflecting the fact that the phase space of 2
particle 1 hole (2p1h) states with this momentum $k$ is considerably larger 
than the corresponding phase space of 2 hole 1 particle states (2h1p).
The scalar and time-like vector part are of similar size and exhibit the
same sign. Using our notation this means that these contributions cancel
each other to a large extent in calculating the expectation value for a
Dirac spinor $u$ representing a nucleon, i.e. a solution of the Dirac
equation at positive energy:
\begin{equation}
\bar{u} (\4v k) \Imag \Sigma (k_0,\3v k ) u (\4v k) = - W^0 (k_0,\3v k ) +
\frac{M^*}{E^*} W^s (k_0,\3v k )+ 
\frac{{\bf \tilde k} \3v k^*}{E^*}
\label{eq:ubwu}
\end{equation}
using the definition
\begin{equation}
{\bf \tilde k} = \3v k  W^v(k_0,\3v k ) \nonumber
\end{equation}
and $E^*$, $M^*$, $ \3v k^*$ as defined in eqs.(\ref{eq:s6}) - (\ref{eq:s8}).
The Dirac spinors $u$ are normalized such that $u^\dagger u = 1$.
As the absolute value of $W^0$
is always larger than the absolute value of $W^s$ at the
same energy, this expectation value, which is roughly proportional
 to $W^s-W^0$, shall be positive at energies below
the Fermi energy and negative above, as it is the case for the imaginary
part of the self-energy calculated within a non-relativistic frame.
The space-component of the vector part, ${\bf \tilde k}$, is significantly
smaller than the other terms. This difference, however, is not as large
as one finds, e.g. in the real part of the self-energy calculated within
the Dirac-Hartree-Fock approach. Similar results are obtained for other
momenta and nuclear densities.

\begin{figure}
\psfig{figure=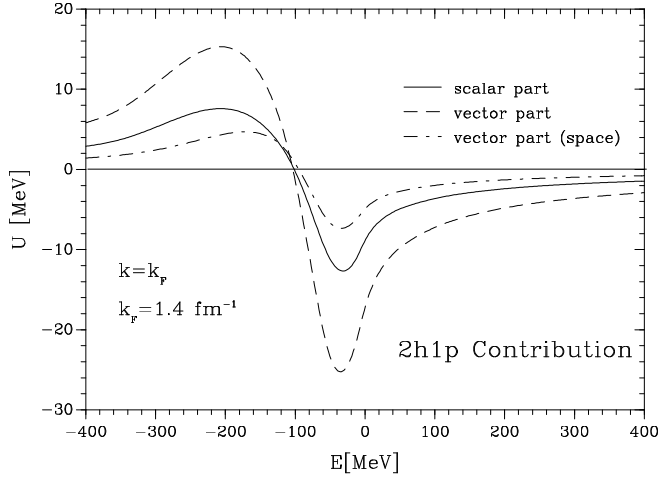,width=9cm}
\caption{Real part of the 2h1p contribution to the self-energy for a fixed
momentum as a function of energy. For further details see Fig.~2.}
\end{figure}

     From the 2h1p contribution to the imaginary part, i.e. the one at
energies below the Fermi energy $E_F^*$ one can determine the
corresponding contribution to the real part by applying the dispersion
relation
\bea
\Delta V^{\alpha}_{2h1p} (\omega ,k) & = &
\Real \Delta \Sigma^{\alpha}_{2h1p} (\omega ,k )\nonumber \\ & = &\frac{P}{\pi}
\int_{-\infty}^{0} d\omega ' \frac{\Imag
\Sigma^{\alpha}_{2h1p} (\omega ' ,k)}{\omega - \omega '} \, ,
\label{eq:disper1}
\eea
where the $P$ is used to indicate the principle value integral and the
index $\alpha$ represents $s, \, 0$, and $v$, referring to the scalar and
vector components of $\Sigma$. Note that here and in the following the energy
variables $\omega$ are redefined such that $\omega =0$ corresponds to the Fermi
energy $k_0 = \epsilon_F$. Typical examples for the real part of the
2h1p contribution to the self-energy are displayed in Fig.~3. The
energy dependence of these terms, can easily
be understood from the energy dependence of the imaginary part, shown in
Fig.~2, and the dispersion relation of eq.(\ref{eq:disper1}). All three
contributions are negative for energies between $-100$ MeV and the Fermi
energy, which are typical energies for quasihole states, as well as for
positive energies, i.e. particle states with energies above the Fermi
energy $\epsilon_F$. As the absolute value
of the time-like vector component $\Real \Delta \Sigma^{0}_{2h1p}$ (dashed
line) is
consistently larger than the corresponding scalar component (solid
line), we obtain a repulsive contribution to the quasiparticle
energy, arising from this 2h1p term from energies starting around 
$-100$ MeV below the Fermi energy up to infinity . This repulsive contribution 
is largest for quasihole states with energies below $\epsilon_F$ and
will decrease for energies above $\epsilon_F$ with increasing energy. It
should be noted that the space-like vector component exhibits a sizable
contribution.

\begin{figure}
\psfig{figure=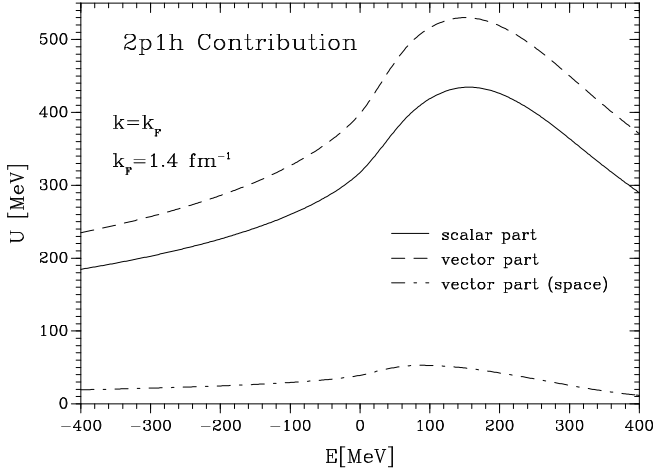,width=9cm}
\caption{Real part of the 2p1h contribution to the self-energy for a fixed
momentum as a function of energy. For further details see Fig.~2.}
\end{figure}

The real part of the 2p1h contribution to the nucleon self-energy can be
calculated from the corresponding imaginary part by a dispersion
relation rather similar to eq.(\ref{eq:disper1})
\be
\Real \Delta \Sigma^{\alpha}_{2p1h} (\omega ,k) = - \frac{P}{\pi}
\int_{0}^{\infty} d\omega ' \frac{\Imag
\Sigma^{\alpha}_{2p1h} (\omega ', k)}{\omega - \omega '} \, .
\label{eq:disper2}
\ee
We already discussed above that
the 2p1h states lead to larger contributions to the imaginary part than 
the 2h1p terms. Therefore it is clear that the 2p1h contributions to the
real part of the self-energy are significantly larger than those
originating from 2h1p terms. This can be seen from Fig.~4, which employs a
scale which is about a factor 10 larger than the one used in Fig.~3 to visualize
the 2h1p contributions. 

In the energy range of interest all three terms are
positive. This means that the 2p1h contribution to the scalar part of the
self-energy $\Sigma^s$ tends to compensate the negative Hartree-Fock
contribution to this term originating mainly from the exchange of two correlated 
pions, which is parameterized in the realistic OBE potentials by means of the
$\sigma$ meson exchange. A similar situation also arises in the case of 
the timelike vector component of the nucleon self-energy $\Sigma^0$: The
negative Hartree-Fock contribution to $\Sigma^0$, which is mainly due to the
exchange of the $\omega$ meson, is compensated to some extent by the 2p1h terms.
This means that the 2p1h terms tend to reduce the Hartree-Fock contributions to
the various terms in the self-energy while the 2h1p corrections yield
contributions to $\Sigma^s$ and $\Sigma^0$ with the same sign as the
Hartree-Fock terms. Similar results are obtained for other momenta $k$.

Since, however, our Hartree-Fock approximation to the self-energy has already
been extracted from a DBHF calculation, we are not allowed to simply add the
real part of 2p1h contribution to the self-energy to the corresponding DBHF
results. This would lead to a double counting of these 2p1h terms. Instead we
use a subtracted dispersion relation defined by
\be
\Delta V^{\alpha}_{2p1h} (\omega ,k) = \Real\Delta\Sigma^{\alpha}_{2p1h} 
(\omega ,k) -\Real\Delta\Sigma^{\alpha}_{2p1h} (\epsilon_{HF}(k) -\epsilon_F,k)
\label{eq:subtract}
\ee
using the real contributions to the various terms $\alpha$ in the self-energy as
defined in (\ref{eq:disper2}). With these definitions we now define a
quasiparticle self-energy, which is real and energy dependent, to be
\be 
V^{\alpha}_{qp} (\omega , k) = \Sigma^{\alpha}_{HF} (k) + \Delta
V^{\alpha}_{2p1h} (\omega ,k) + \Delta V^{\alpha}_{2h1p} (\omega
,k)\label{eq:sigqp}
\ee
and we can determine a quasiparticle spinor 
\be
u_{qp}(\3v k) = \sqrt{\frac{E^*_{qp}+M^*_{qp}}{2E^*_{qp}}}\left(\begin{array}{c}
 1 \\ \frac{\vec{\sigma}\cdot \3v k^*_{qp}}{E^*_{qp}+M^*_{qp}}\end{array}\right)
\label{eq:uqp}
\ee 
with
\begin{eqnarray}
M^*_{qp} & = & M + V^s_{qp} (\omega_{qp},k)\nonumber\\
\3v k^*_{qp} & = & \3v k (1 + V^v_{qp} (\omega_{qp},k)\nonumber\\
E^*_{qp} & = & \sqrt{(\3v k^*_{qp})^2 + (M^*_{qp})^2}\nonumber\\
\omega_{qp} & = & E^*_{qp} - V^0_{qp} (\omega_{qp},k) - \epsilon_F\,.
\label{eq:mstarqp}
\end{eqnarray}
This means that this spinor is an eigenstate of the Dirac equation
\be
\left[(1+V^v_{qp}) \vec\alpha \cdot \3v k + \gamma^0(M+V^s_{qp})
-V^0_{qp}\right] u_{qp} = \epsilon_{qp} u_{qp} \ ,
\label{eq:diraceq} 
\ee
which uses the
self-energy $V_{qp}$ of (\ref{eq:sigqp}) calculated at the energy $\omega$
which corresponds to the quasiparticle energy defined by
\be
\epsilon_{qp} (k) = E^*_{qp} -\Sigma^0_{qp} (\omega_{qp},k) \ .
\label{eq:epsqp}
\ee
If for the moment we ignore the 2h1p
contribution,
$\Delta V_{2h1p}^\alpha$, to the quasiparticle self-energy in
(\ref{eq:sigqp}), the subtraction defined in (\ref{eq:subtract}) 
ensures that the
$\Delta V_{2p1h}^\alpha$ terms vanish on-shell and the quasiparticle energy
coincides with the $HF$ energy, thus avoiding double counting.
However, when the contribution of the 2h1p terms is
taken into account, the self-consistent definition of the energy variable
$\omega_{qp}$ gives rise to a non-vanishing correction due to energy
dependence of the 2p1h terms.

\begin{figure}
\psfig{figure=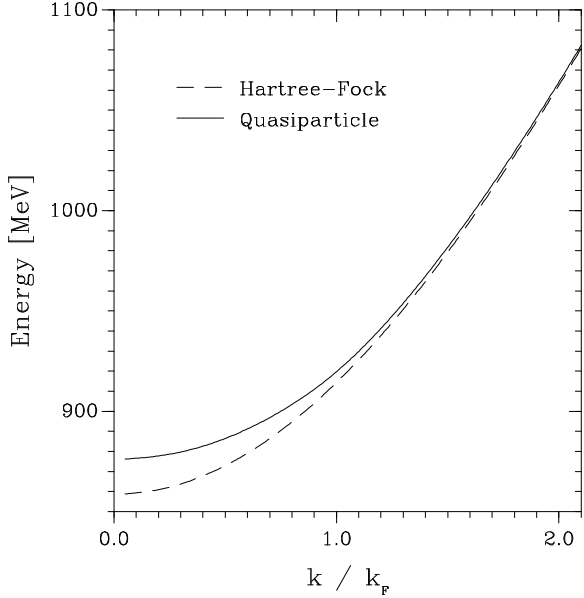,width=8cm}
\caption{Results for the quasiparticle energy, calculated according to eq. 
(\protect\ref{eq:epsqp}), are compared to the correponding energies calculated
in the HF approximation for the self-energy (dashed line)}
\end{figure}

Results for the quasiparticle energy 
are displayed in Fig.~5 for various momenta $k$. It can bee seen that the
inclusion of the 2h1p terms yields a significant reduction in the quasiparticle
energy $\epsilon_{qp}$ as compared to the corresponding HF result. This
reduction is as large as 17.5 MeV for momenta close to zero and reduces to
values around 5.3 MeV for $k=k_F$ and becomes negligibly small around $k=2k_F$.
This repulsive effect of the 2h1p terms in the quasiparticle energy has been 
discussed already in context with Fig.~3. Such a large effect for the
quasiparticle energy may indicate that the 2h1p terms may as well have some
effect in the calculation of the total energy. For the calculation of the total
energy, however, it is not sufficient to evaluate the quasiparticle energy, but
one would need the whole spectral distribution of hole strength for states with
momenta below and above the Fermi momentum\cite{wimrev}.

It is one aim of our study to explore the effects of the Dirac structure of the
nucleon self-energy, calculated beyond the mean field approximation, on the
complex optical potential for nucleon-nucleus scattering. 
The Dirac equation reads now:
\be
\left[(1+\Sigma^v_{qp}) \vec\alpha \cdot \3v k + \gamma^0(M+\Sigma^s_{qp})
-\Sigma^0_{qp}\right] u_{qp} = \tilde\epsilon_{qp} u_{qp}
\label{eq:diraccomp}
\ee
where the full complex self-energy is used. Real $\tilde\epsilon_{qp}$
solutions of
eq. (\ref{eq:diraccomp}) involve complex values of $\3v k$. We take
as approximate solutions of eq. (\ref{eq:diraccomp}) the values 
$\epsilon_{qp}$ determined from eq. (\ref{eq:epsqp}).
It is convenient to rewrite eq. (\ref{eq:diraccomp}) into a form 
which only contains an effective scalar ${\cal V}^s$ and
vector potential ${\cal V}^0$\cite{mitsumo}
$$
\left[ \vec\alpha \cdot \3v k + \gamma^0(M+{\cal V}^s)
-{\cal V}^0\right] u_{qp} = \epsilon_{qp} u_{qp}
$$ where
\bea
{\cal V}^s & = & \frac{\Sigma^s_{qp}-M\Sigma^v_{qp}}{1+\Sigma^v_{qp}}\,,
\nonumber\\
{\cal V}^0 & = & \frac{\Sigma^0_{qp}-\epsilon_{qp}\Sigma^v_{qp}}
{1+\Sigma^v_{qp}}\,.\label{eq:renorsig}
\eea
The self-energy terms $\Sigma^{\alpha}_{qp}$ have been calculated in the
quasiparticle approach of (\ref{eq:sigqp}) using the self-consistent relation
between the three-momentum and the energy as defined in (\ref{eq:mstarqp}).
This Dirac equation can be transformed into a Schr\"odinger-type equation 
for the
large component of the Dirac spinor, leading to a complex and energy-dependent 
Schr\"odinger equivalent potential of the form
\be
{\cal U} = {\cal V}^s - \frac{\epsilon_{qp}}{M}{\cal V}^0 + 
\frac{1}{2 M}\left[({\cal V}^s)^2-
({\cal V}^0)^2\right]\,.
\label{eq:schroed}
\ee

\begin{figure}
\psfig{figure=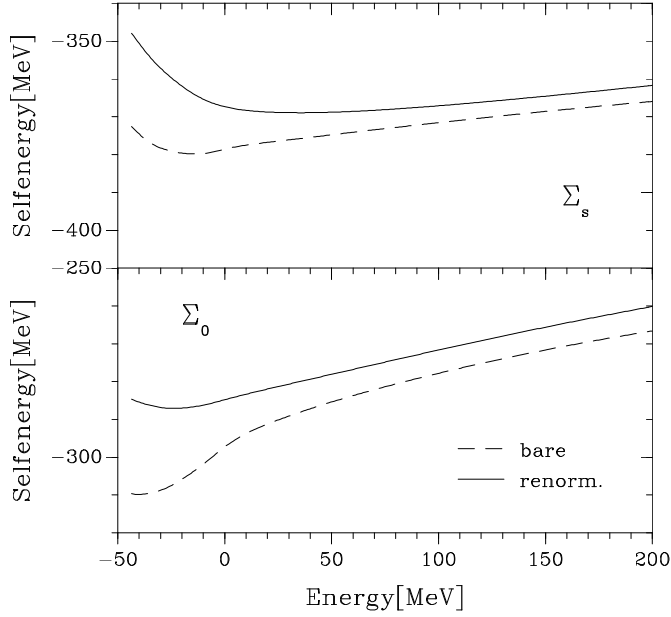,width=9cm}
\caption{The real parts of the bare scalar and vector components 
($\Sigma^s_{qp}$ and
$\Sigma^0_{qp}$), represented by the dashed line, are compared to the 
real parts of
${\cal V}^s$ and ${\cal V}^0$, renormalized according to eq. 
(\protect\ref{eq:renorsig}). 
}
\end{figure}

Results for the real part of the
renormalized components ${\cal V}^s$ and ${\cal V}^0$ are
displayed in Fig.~6 as a function of the energy variable $\omega_{qp}$, i.e.~the
quasiparticle energy $\epsilon_{qp}$ normalized such that the Fermi energy
occurs at zero. For a comparison we also show the unrenormalized components
$V^{\alpha}_{qp}$. The difference between the solid and dashed line is a
measure for the importance of the space-like vector component $\Sigma^v$ of the
self-energy. We find that this space like components yield a slight reduction of
the absolute values for Real${\cal V}$, which is of the order of 
3 percent. Larger
effects only occur at negative energies, below the Fermi energy,
where the 2h1p contribution gets more important.

\begin{figure}
\psfig{figure=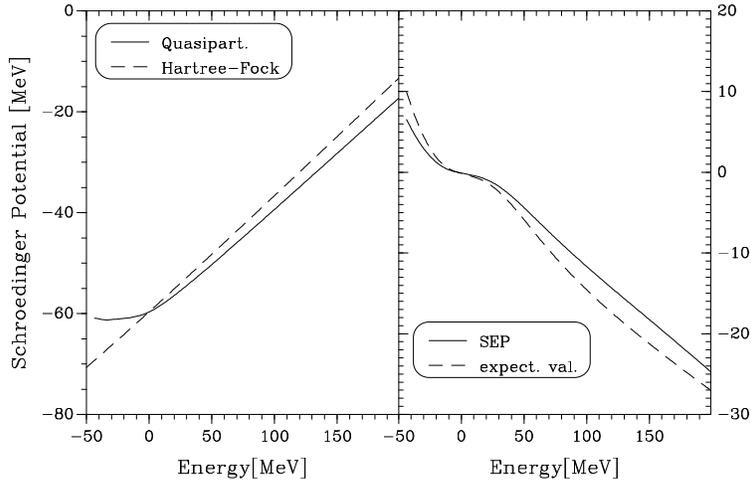,width=10cm}
\caption{Energy dependence of the real (left) and imaginary part (right)
of the Schr\"odinger equivalent potential. The result for
the real part is compared to the corresponding prediction obtained in the HF
approximation. The dashed curve in the right part of the figure exhibits the
expectation value of $W$ calculated according to (\protect \ref{eq:wexpt}).
}
\end{figure}

This figure also demonstrates that the energy-dependence of the various Dirac
components in the real part of the self-energy remains weak, again with the
exception of energies below $\epsilon_F$. Also the deviations from the
Hartree-Fock result, -375 MeV and -290 MeV for the scalar and vector parts,
respectively, are not very pronounced. This can also be seen from the left part
of Fig.~7, exhibiting the real part of the
Schr\"odinger equivalent potential ${\cal U}$ defined
in (\ref{eq:schroed}). The results derived from the quasiparticle approximation
(solid line) including the effects of the energy-dependence in the 2p1h and 
2h1p terms, exhibit a dependence on the quasiparticle energy, which is very
similar to the one derived from the HF approximation. Therefore one may conclude
that the energy-dependence in the depth of central, Woods-Saxon type, optical
potentials, used to describe nucleon-nucleus scattering is mainly due to the
relativistic structure of the underlying Hartree-Fock 
self-energy to be used in a Dirac
equation. Dispersive effects due to the consideration of 2p1h and 2h1p terms
also lead to an energy dependence, which, however, is much smaller.
These dispersive corrections tend to make the potential slightly
more attractive at higher energies. It should be mentioned, however, that an
energy dependence of the central Schr\"odinger potential similar to the one
obtained here within the relativistic scheme can also be obtained within
a non-relativistic Hartree-Fock due to large non-localities in the Hartree-Fock
potential\cite{kleinm}.

\begin{figure}
\psfig{figure=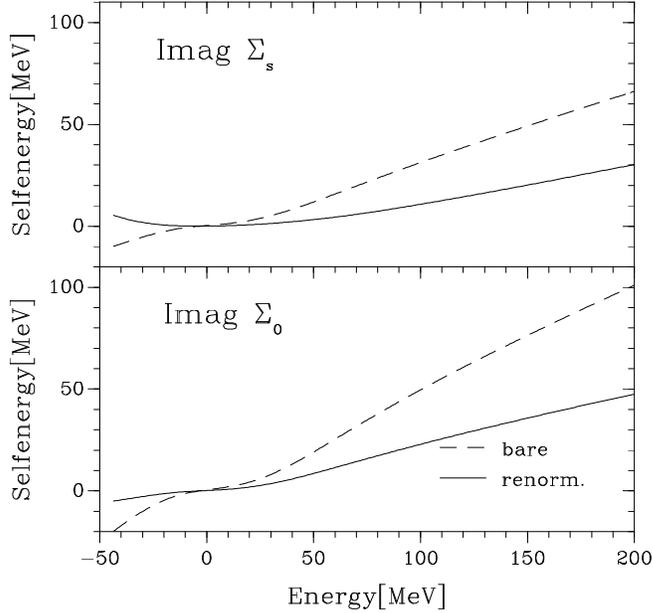,width=9cm}
\caption{Bare and renormalized scalar and vector components of the imaginary
part of the quasiparticle self-energy. For further details see Fig.~6.}
\end{figure}

An analysis rather similar to the one just outlined for the real components of
the self-energy can also be performed for the imaginary parts. 
Results for the renormalized Dirac components ${\cal W}^{\alpha}=
{\rm Imag} {\cal V}^{\alpha}$ ($\alpha=s$ and 0) are
presented in Fig.~8 and compared to the bare scalar- and vector imaginary
components. As we discussed already in the beginning of this section, the
imaginary part of the bare
components are negative for energies below the Fermi energy (2h1p), 
zero around $\omega=0$ ($\epsilon_F$)
and positive for positive energies (2p1h). The renormalizing
effects of the space like vector components are much more important for the
imaginary part than for the real part. This can be deduced from the differences
between ${\cal W}^{\alpha}$ and ${ W}^{\alpha}$: the bare components are as
large as twice the renormalized quantities. 

The imaginary part of the Schr\"odinger optical
potential [eq. (\ref{eq:schroed})] is shown on the right of Fig.~7. 
An alternative way of
evaluating this imaginary part would be to calculate the expectation value of
the Dirac operator
\be 
\bar u_{qp}\left[W^s - \gamma^0 W^0 + \mbox{\boldmath$\gamma$}\cdot{\3v
k}W^v\right] u_{qp}\label{eq:wexpt}
\ee
using the quasiparticle Dirac spinors of (\ref{eq:uqp}). These expectation 
values lead to the dashed line on the right part of Fig.~7. 
We see that the results for the imaginary
part of the Schr\"odinger potential are rather insensitive on the way of
calculation.

\begin{figure}
\psfig{figure=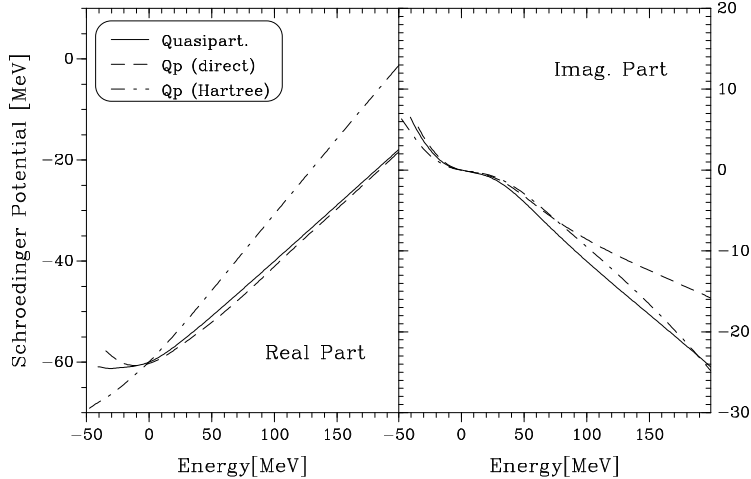,width=10cm}
\caption{Real and Imaginary parts of the Schr\"odinger equivalent potential,
calculated using the full model (solid lines),
ignoring exchange diagrams in the evaluation of the terms of second order
(dashed line), and using the Hartree approximation also for the lowest order
term (dashed-dotted line)}
\end{figure}

In the calculation of the imaginary components of the self-energy discussed 
in section 2 and consequently also in the corresponding real components, we
always determined the Direct- and Exchange-contributions (see eqs. (\ref{eq:Md})
and (\ref{eq:Me})). In order to explore the importance of the 2p1h and 2h1p
exchange terms we compare in Fig.~9 the results obtained for the real and
imaginary part of the Schr\"odinger equivalent potential with (solid lines) and
without (dashed lines) inclusion of the exchange terms in the 2p1h and 
2h1p parts of the
self-energy. One finds that the effects of the exchange terms on the real part
are rather small typically around 1 MeV. However, the effects are significantly
larger for the imaginary part where the difference gets as large as 30 percent
of the total result. This can be understood from the fact that the real part of
the self-energy is dominated by the Hartree-Fock contribution, which is
identical in these two approaches.

If one ignores the exchange terms in calculating the 2p1h and 2h1p terms, one
may consider it more consistent to ignore the exchange term 
also in the leading
contribution and replace the Hartree-Fock approximation by
the Hartree-approach. In ref.\cite{fritz} effective meson-nucleon coupling 
constants were determined to reproduce the DBHF results within a Dirac-Hartree
model. The resulting coupling constants are a bit larger than those derived from
the Dirac-Hartree-Fock analysis. If we use these Hartree-coupling constants and
ignore the effects of exchange terms in the leading term as well as in the 2p1h
and 2h1p terms, one arrives at a Schr\"odinger equivalent potential displayed by
the dashed-dotted lines in Fig.~9. The differences to the solid lines get now
quite pronounced for the real part. This can be traced back to the fact
that in the Dirac-Hartree approximation $\Sigma^s$ and $\Sigma^0$ are
independent on the momentum or energy of the state and $\Sigma^v$ is identical
to zero. The results for the imaginary part obtained in this approach, however,
are rather close to those evaluated with inclusion of the echange contributions.

\begin{figure}
\psfig{figure=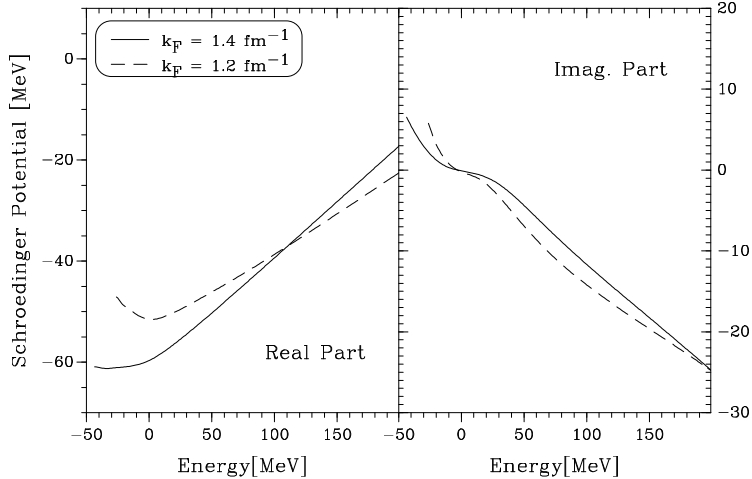,width=10cm}
\caption{Real and Imaginary parts of the Schr\"odinger equivalent potential
calculated for nuclear matter with a Fermi momentum of $k_F$ = 1.4 fm$^{-1}$
(solid line) are compared to those obtained for $k_F$ = 1.2 fm$^{-1}$.}
\end{figure}

It is also interesting to investigate the density dependence of the
self-energy in order to extend these calculations to finite nuclei. 
As an example we present some results
obtained for nuclear matter with a Fermi momentum $k_F$ = 1.2 fm$^{-1}$ in
Fig.~10. The results obtained at these various densities, either for the
self-energy keeping track of the Dirac structure, or using the Schr\"odinger
equivalent potentials derived from these components, may than be used in a local
density approximation for a prediction of nucleon-nucleus scattering. Such
investigations are in progress.

\section{Conclusions}
The relativistic structure of the self-energy for a nucleon in nuclear matter is
investigated by including all irreducible terms of first and second order in the
residual interaction. For the NN interaction a parameterization of the
Dirac-Brueckner-Hartree-Fock (DBHF)
$G$-matrix in terms of the exchange of effective
scalar and vector mesons has been used. The 2p1h
and  2h1p contributions to the imaginary part of the
self-energy ar
e evaluated keeping track of all direct and exchange terms.
The corresponding 2p1h and 2h1p contributions to the real part are derived from
these imaginary components by means of dispersion relations. A subtracted
dispersion relation must be used for the 2p1h term to avoid double-counting with
the $G$-matrix underlying the DBHF approach on which these studies are based.

The inclusion of 2h1p diagrams in the evaluation of the real part of the
self-energy yields a non-negligible modification of the scalar and vector
components in particular for states with momenta below the Fermi momentum. 
Also the value of the quasiparticle energy is increased by a value 
 as large as 17.5 MeV for momenta close to zero and to values around 5.3 MeV 
for $k=k_F$. Such a large effect for the
quasiparticle energy may indicate that the 2h1p terms should have some
effect in the calculation of the total energy. 

The calculated self-energy can also be transformed into a Schr\"odinger
equivalent optical potential, to be used in the study of nucleon-nucleus
scattering. Exchange diagrams are non-negligible in the evaluation of the
imaginary part. The energy- or momentum-dependence of the central component of
the real potential, however, is dominated by the effects of the 
Dirac-Hartree-Fock
contribution. The 2p1h and 2h1p terms give rise to a more attractive
SEP at positive enrgies and introduce an additional energy dependence
which is very weak.

This project has
been supported by the Spanish research grant DGICYT, PB95-1249, and by the EC
contract CHRX-CT93-0323. One of us (H.M.) is pleased to acknowledge the warm
hospitality at the Facultat de F\'\i sica, Universitat de Barcelona, and the
support from the program of visiting Professor at this University.

\end{document}